\documentclass[12pt]{iopart}
\usepackage{iopams}
\usepackage{setstack}

\usepackage{graphicx}
\usepackage{dcolumn}
\usepackage{bm}
\usepackage{float}
\usepackage[english]{babel} 
\usepackage[T1]{fontenc}

\usepackage[numbers,sort&compress]{natbib}

\usepackage[usenames,dvipsnames]{xcolor}
\usepackage[colorlinks=true,citecolor=Cerulean,linkcolor=RubineRed,urlcolor=Cerulean]{hyperref}

\newcommand{\ket}[1]{|{#1}\rangle}

\begin{document}

\title{The Mean-Field Bose Glass in Quasicrystalline Systems}
\author{Dean Johnstone$^1$, Patrik \"{O}hberg$^1$ and Callum W. Duncan$^{2,3}$}
\address{$^1$ SUPA, Institute of Photonics and Quantum Sciences,
	Heriot-Watt University, Edinburgh, EH14 4AS, UK}
\address{$^2$ Department of Physics, SUPA and University of Strathclyde, Glasgow G4 0NG, United Kingdom}
\address{$^3$ Max-Planck-Institut f\"{u}r Physik komplexer Systeme, D-01187 Dresden, Germany}
\ead{dj79@hw.ac.uk, p.ohberg@hw.ac.uk, callum.duncan@strath.ac.uk}
\date{\today}

\begin{abstract}
We confirm the presence of a mean-field Bose glass in 2D quasicrystalline Bose-Hubbard models. We focus on two models where the aperiodic component is present in different parts of the problem. First, we consider a 2D generalisation of the Aubry-Andr\'e model, where the lattice geometry is that of a square with a quasiperiodic onsite potential. Second, we consider the randomly disordered vertex model, which takes aperiodic tilings with non-crystalline rotational symmetries, and forms lattices from the vertices and lengths of the tiles. For the disordered vertex models, the mean-field Bose glass forms across large ranges of the chemical potential, and we observe no significant differences from the case of a square lattice with uniform random disorder. Small variations in the critical points in the presence of random disorder between quasicrystalline and crystalline lattice geometries can be accounted for by the varying coordination number and the different rotational symmetries present. In the 2D Aubry-Andr\'e model, substantial differences are observed from the usual phase diagrams of crystalline disordered systems.  We show that weak modulation lines can be predicted from the underlying potential and may stabilise or suppress the mean-field Bose glass in certain regimes. This results in a lobe-like structure for the mean-field Bose glass in the 2D Aubry-Andr\'e model, which is significantly different from the case of random disorder. Together, the two quasicrystalline models studied in this work show that the mean-field Bose glass phase is present, as expected for 2D quasiperiodic models. However, a quasicrystalline geometry is not sufficient to result in differences from crystalline realisations of the Bose glass, whereas a quasiperiodic form of disorder can result in different physics, as we observe in the 2D Aubry-Andr\'e model.
\end{abstract}


\section{Introduction}
The Bose Glass (BG) is a well-known phase that arises in disordered Bose-Hubbard models \cite{Fisher1989,PhysRevB.80.214519,PhysRevLett.99.050403,PhysRevB.73.174523,PhysRevB.55.R11981,PhysRevB.53.2691},  which separates a direct Mott Insulator (MI) to Superfluid (SF) transition \cite{Fisher1989,PhysRevLett.103.140402}.  The BG contains small regions of a SF nature, but it is a macroscopic insulator; with no transport exhibited across the full system.  The phase is characterised by a finite compressibility,  the lack of a gap, and an infinite SF susceptibility \cite{Fisher1989}. This means its properties are different from that of the SF or MI states present in the disorder free system.  In this work, by using common mean-field methods and a percolation analysis, we will confirm the existence of a mean-field BG phase in quasicrystalline systems.

Over recent years, quasiperiodic, or quasicrystalline, systems have gained significant interest as they represent the middle ground between correlated, periodic order and uncorrelated, random disorder \cite{PhysRevB.87.134202,schreiber2015observation,duncan2017a,PhysRevB.100.104203,mace2019many}. Unlike their periodic counterparts, quasicrystals do not possess translational invariance \cite{senechal1996quasicrystals,Mas_kov__1998}. Despite this, they still have long-range order, which can be seen from their self-similarity.   Quasiperiodic models are also known to host intriguing dynamics \cite{PhysRevB.33.3837,PhysRevB.99.224204,PhysRevB.96.180204,PhysRevB.97.174206}, multi-fractal properties \cite{xu1987fractal,PhysRevB.99.224204,PhysRevB.99.155121}, and the contentious possibility to stabilise many-body localised phases in 2D, due to the absence of conventional rare regions seen in disordered systems \cite{PhysRevB.87.134202,PhysRevB.100.104203,mace2019many,agarwal2017rare,PhysRevB.93.134206}. 

A ubiquitous example of a quasiperiodic system is the 1D Aubry-Andr\'e (AA) model, where it is known that the underlying self-duality separates regimes where states are either localised or extended \cite{aubry1980analyticity}, which prevents the formation of a single-particle mobility edge in the spectrum \cite{mott1987mobility}. The expansion of the AA model itself to a 2D square lattice has also recently been gaining interest \cite{PhysRevLett.116.140401,PhysRevB.99.054211,PhysRevB.101.014205} and shares many of the same properties as that of the 1D AA model. Another method to model quasicrystals is by considering a vertex model of an aperiodic tiling \cite{Odagaki1986,Kohmoto1986,Arai1988,Tsunetsugu1991,Repetowicz1998}. In a vertex model, edges of tiles are taken to be bonds and vertices of the tiling pattern are taken to be lattice sites. These kinds of quasicrystalline lattices have been used extensively in prior works to study single-particle physics on aperiodic systems, including their transport \cite{PhysRevLett.71.4166,PhysRevLett.70.3915}, and topological properties \cite{PhysRevB.91.085125,duncan2019topological,PhysRevB.94.205437}.  

The concept of disorder in quantum physics represents the absence of some intrinsic symmetry or structure to a particular system. In this context, impurities and defects to a crystalline material can dramatically change observable properties.  In the single-particle picture,  it is known that random variations can result in Anderson localisation \cite{PhysRev.109.1492,RevModPhys.50.191}.  When coupled with interactions, competing energy scales can introduce further novel properties in the form of exotic phase transitions \cite{Ruiz_Lorenzo_1997,fisher1990quantum,vojta2010quantum,vojta2013phases},  many-body localisation \cite{PhysRevB.21.2366,PhysRevLett.78.2803,PhysRevB.82.174411,nandkishore2015many}, and glass states \cite{Fisher1989,chakravarty1999wigner,schmitteckert1998fermi,freedman1977theory}.  Hubbard models describing random parameters were first introduced in the context of disordered fermionic and spin systems \cite{PhysRevB.37.325,PhysRevLett.10.159,PhysRev.115.2}, but have also been studied in the bosonic scenario. In particular, the disordered Bose-Hubbard model has been used to describe superfluid helium in porous media \cite{Reppy1992SuperfluidHI}, granular superconductors \cite{PhysRevB.64.054515}, Josephson junctions \cite{PhysRevB.67.014518}, and ultracold atoms in speckle potentials \cite{PhysRevLett.91.080403,Billy2008}. 

In this work, we will confirm the presence of a mean-field BG in two distinct quasicrystalline models. First, we will consider a quasiperiodic potential supported on a square lattice, in the form of a 2D AA model. This model will show that the mean-field BG can indeed exhibit self-similar properties of the underlying potential. Second, we will consider quasicrystalline lattices generated from vertex models of aperiodic tilings. In these models -- which we will label as disordered vertex models -- we have a quasicrystalline lattice with a fully disordered potential. From these two models, we can probe the influence of the quasiperiodic nature arising from either the potential or lattice structure independently.  We will show that the mean-field BG is only substantially changed when the quasiperiodic component of the model is present in the potential,  with a quasicrystalline form of the lattice only shifting transition points in a minor way.

\section{Quasicrystalline Models} \label{sc_2}

For all models, we will consider an inhomogeneous Bose-Hubbard Hamiltonian,  with nearest-neighbour tunnelling, given by
\begin{equation}	\label{eq_bhm}
\hat{H} = \frac{U}{2} \sum_i^N \hat{n}_i (\hat{n}_i -1) + \sum_i^N (\epsilon_i - \mu) \hat{n}_i - J\sum_{\langle i,j\rangle} \left(\hat{b}^\dagger_i \hat{b}_j + H.c. \right) ,
\end{equation}
where $N$ is the total number of lattice sites, $U$ is the onsite interaction strength, $\epsilon_i$ is an onsite energy offset, $J$ is the tunnelling coefficient, $\langle i,j\rangle$ denotes the sum over nearest-neighbours,  and $\mu$ is the chemical potential.  The operators $ \hat{b}_i \, (\hat{b}^\dagger_i) $ are the individual bosonic annihilation (creation) operators with $\hat{n}_i$ being the number operator. 

\subsection{Interacting 2D Aubry-Andr\'e Model}
\label{sec:InterAA}

\begin{figure}
	\centering,
	\makebox[0pt]{\includegraphics[width=0.98\linewidth]{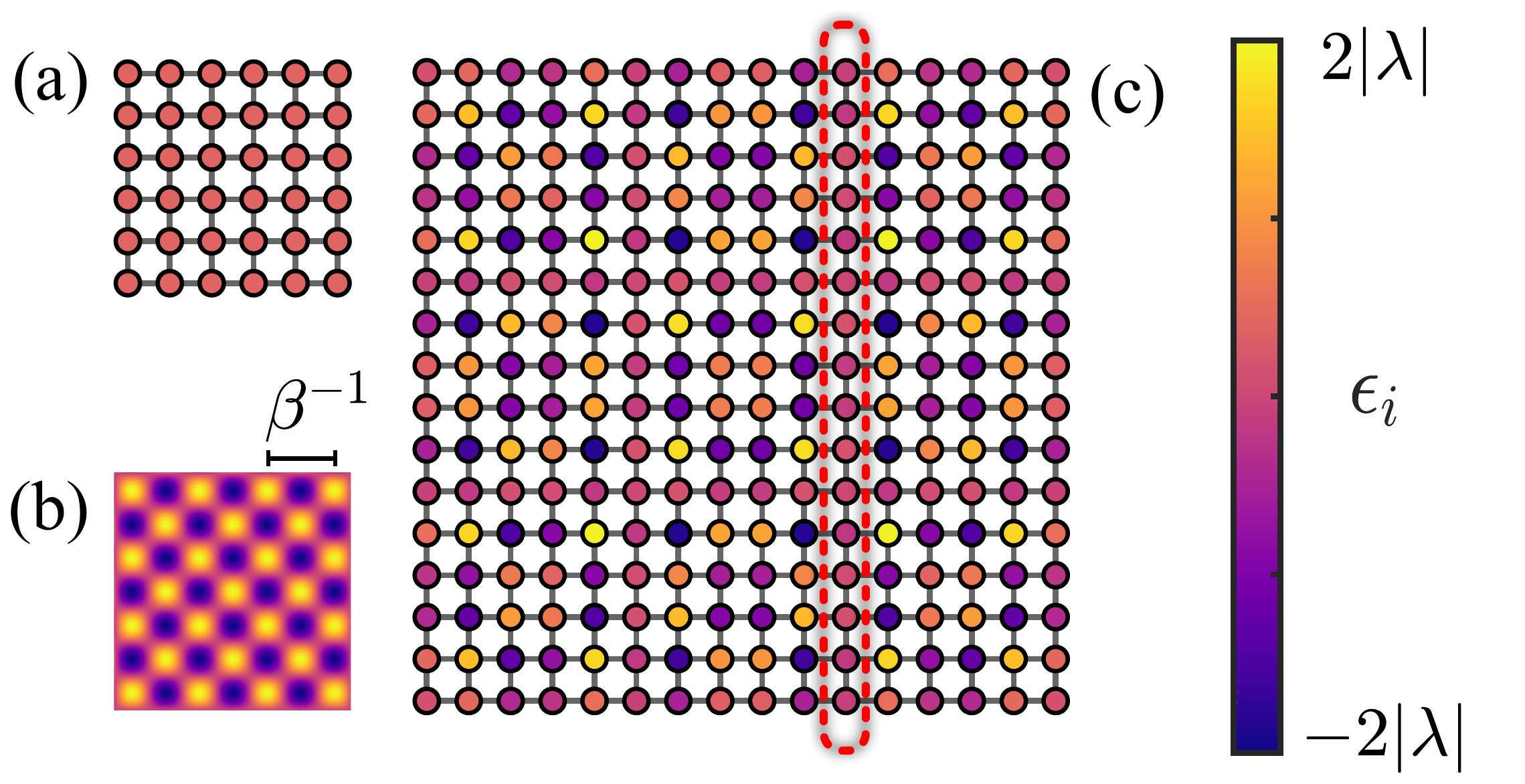}}
	\caption{Visualisation of the 2D Aubry-Andr\'e model on a finite lattice. Here, we show (a) the underlying square lattice with unit spacing and (b) the continuum limit of Eq. \eref{eq_aaEps} with period $\beta^{-1}$ in the same spatial interval. (c) The 2D AA model can be formed from a combination of (a) and (b) on a larger lattice. The dashed (red) box highlights a particular line of weak modulation for the 2D AA potential.}
	\label{figure_x1}
\end{figure}

The first model we will look at is an interacting version of the AA model generalised to 2D. We consider a 2D square lattice with unit lattice spacing and assume that the $\epsilon_i$ of Eq. \eref{eq_bhm} are distributed according to the AA quasiperiodic potential
\begin{equation} 	\label{eq_aaEps}
\epsilon_i = -\lambda \Big( \cos[2 \pi \beta (x_i+ y_i)] + \cos[2 \pi \beta (x_i - y_i)] \Big),
\end{equation}
with $\lambda$ denoting the modulation strength, $\beta$ is the wavenumber, and $x_i \, (y_i)$ are the position indices.  A phase can be incorporated into the distribution of $\epsilon_i$ and averaged over to allow for a direct analogy of the quasiperiodic system with random disorder \cite{roati2008anderson,PhysRevB.96.054202,PhysRevA.78.023628,PhysRevLett.98.130404,PhysRevLett.91.080403}. However, we are interested in Eq. \eref{eq_aaEps} for its quasiperiodic properties, and therefore the introduction of a phase is irrelevant.  Note,  alternative extensions of the AA model to 2D have been considered, including coupled 1D chains \cite{PhysRevLett.116.140401} which can exhibit mobility edges \cite{PhysRevB.99.054211}.

For illustrative purposes, we plot a visualisation of the AA potential in Fig. \ref{figure_x1} for a small lattice. As we can clearly see in Fig. \ref{figure_x1}(c), the discretised potential can form a pattern, with weakly modulated lines appearing throughout the lattice. In a recent study within the single-particle picture, it was shown that these weakly modulated lines prevent a mobility edge from separating localised/extended states and can support ballistic transport \cite{PhysRevB.101.014205}. As we will see later, the weakly modulated lines will also play an important role in the formation and structure of phases in the mean-field scenario.

\subsection{Disordered Vertex Model}
\label{sec:DisVert}

\begin{figure}
	\centering,
	\makebox[0pt]{\includegraphics[width=0.8\linewidth]{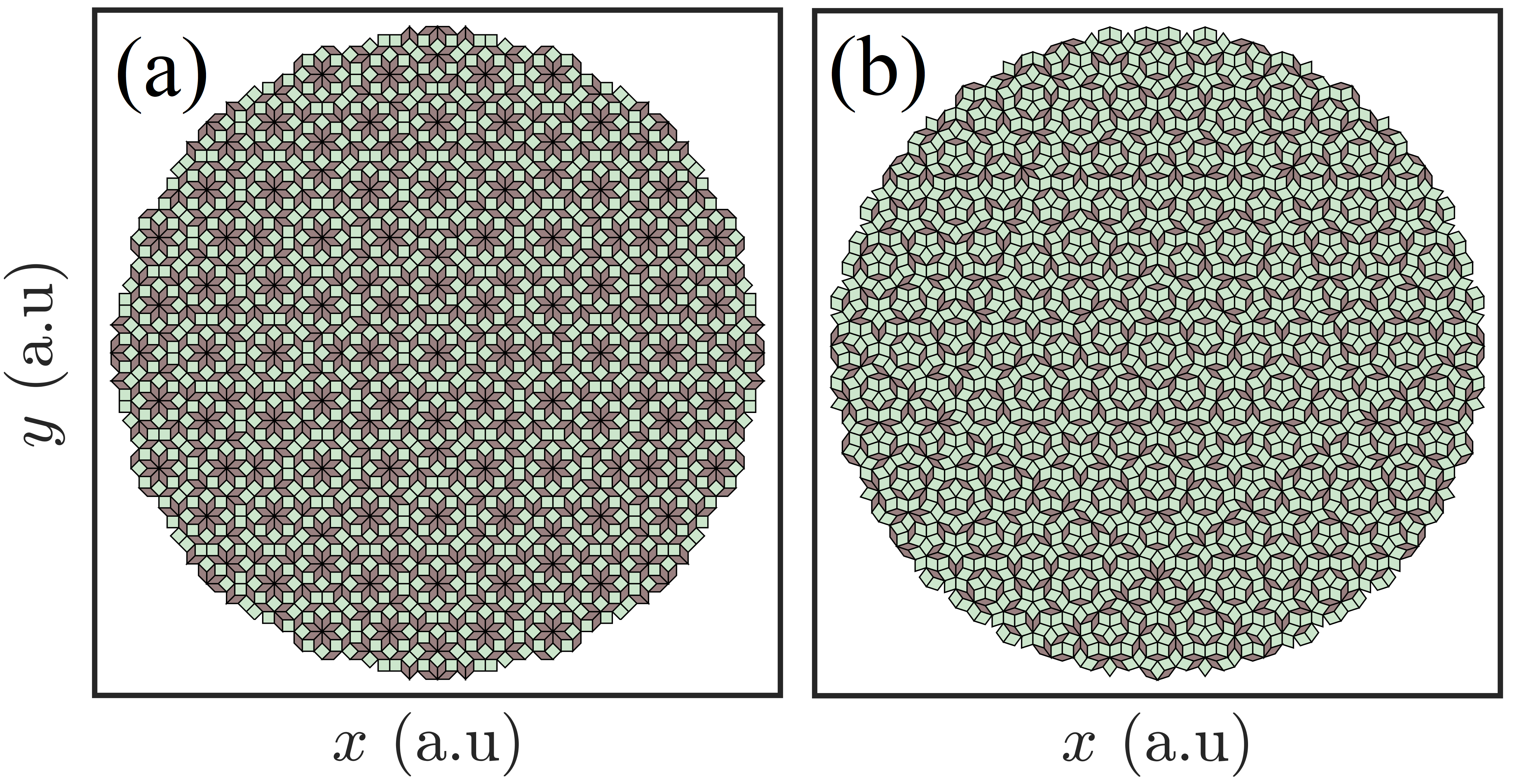}}
	\caption{Tilings generated from the cut-and-project method, showing (a) the Ammann-Beenker tiling and (b) Moore-Penrose tiling. In both cases, the tilings are composed of two rhombic unit cells (shaded and unshaded rhombi on the tilings) that fill all of space continuously. For practical purposes, a circular cut-off is set in real space to limit the total number of tiles / lattice sites and preserve rotational symmetry with respect to the origin. The corresponding vertex model of both tilings has (a) 3041 and (b) 3056 lattice sites respectively.}
	\label{figure_x2}
\end{figure}

We will also consider the Bose-Hubbard model with random disorder on lattices defined via the vertex model of aperiodic tillings. The lattices we work with are generated from rotationally symmetric quasicrystalline plane tilings \cite{strandburg1991quasicrystals,Mas_kov__1998}.  Note, the strict group theory definition of a lattice is that of a repeating set of points, and in general the models we study in this work, and any model with open boundary conditions, are labelled as graphs. However, the use of the term lattice to describe models with open boundary conditions and quasiperiodic sets of points is now ubiquitous in physics and, hence, we will label these sets of points as lattices throughout this work.
 
\begin{figure}
	\centering,
	\makebox[0pt]{\includegraphics[width=0.8\linewidth]{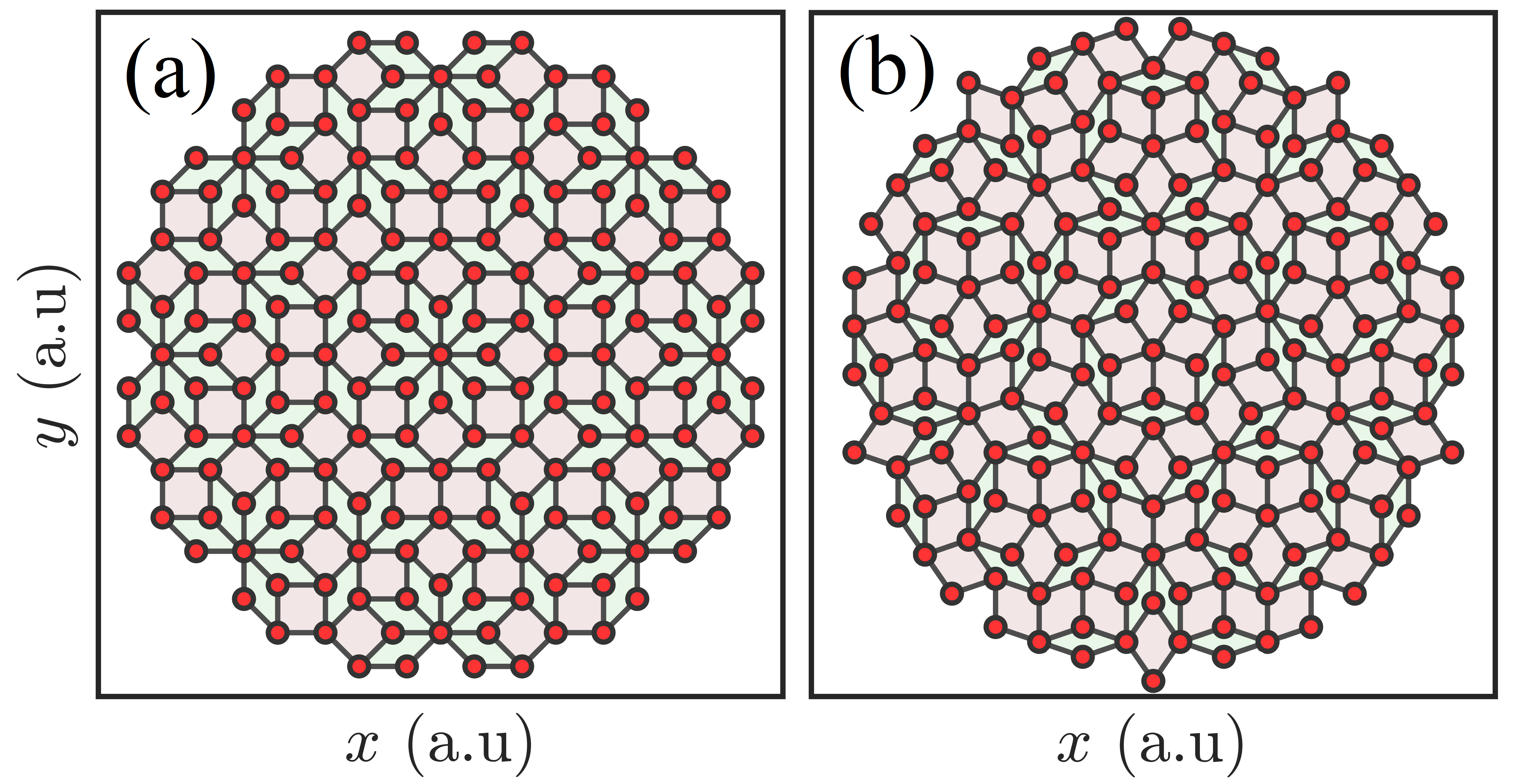}}
	\caption{Smaller tilings generated from the cut-and-project method, showing the (a) Ammann-Beenker tiling and (b) Moore-Penrose tiling, with the vertex model for each on the foreground. Lattice sites correspond to the intersection of tile edges, where the edges are bonds between sites. The reduced lattices contain (a) 192 and (b) 196 sites.}
	\label{figure_x3}
\end{figure}

The link between aperiodic tilings and quasicrystals was further emphasised from Shechtman's original discovery \cite{PhysRevLett.53.1951,PhysRevB.34.3345,PhysRevB.55.3520}, allowing for quasicrystals to be interpreted as canonical projections of higher dimensional crystalline lattices \cite{PhysRevB.39.10519,lagarias1996meyer}.  Cut-and-project sets, in particular, provide a useful framework that we will use to generate quasicrystalline lattices. For further details on this approach, we refer the reader to \ref{sc_QC}. By using the cut-and-project method, we generate the Ammann–Beenker (AB) and Moore-Penrose (MP) tilings from 4D and 5D hypercubic lattices respectively. In Fig. \ref{figure_x2}, we show the real-space structure of the tilings, which clearly illustrates the self-similarity and aperiodicity of quasicrystals over large length scales.  In a vertex model, we take edges of tiles to be bonds and vertices to be lattice sites.  To better visualise the lattice geometry, smaller tilings and their respective vertex models are considered in Fig. \ref{figure_x3}. Both plots also show the 8-fold and 5-fold rotational symmetry of the AB and MP tilings respectively.

In the disordered vertex model, the potential will be random and will contain no self-similarity, unlike the 2D AA model already defined. In particular, we assume that the $\epsilon_i$ in Eq. \eref{eq_bhm} are drawn from a uniform random distribution, with 
\begin{equation}
\epsilon_i \in [-\Delta/2 , \Delta/2], 
\end{equation}
where $\Delta$ is the disorder strength. However, the underlying lattice geometry is aperiodic, with local disorder in the coordination number and a long-range self-similar structure.


\section{Gutzwiller Mean-Field} \label{sc_3}
To study the phases that may arise in these models, we employ a self-consistent mean-field approach. This is done by using the well-known Gutzwiller ansatz \cite{PhysRevLett.10.159,Gutzwiller1965,Rokhsar1991,Krauth1992} for the many-body wavefunction, which takes the form
\begin{equation} \label{eq_gpw}
| \Psi \rangle = \prod\limits_{i}^{N} \sum\limits_{n=0}^z f_n^{(i)} \ket{n_i},
\end{equation}
where $z$ is the maximum number of particles per site, $f_n^{i}$ is the amplitude for $n$ atoms in site $i$ and $\ket{n_i}$ is the corresponding number state. By separating the many-body wavefunction into onsite products, we neglect quantum correlations, which is valid in the limit for small $J/U$ and a large number of nearest-neighbours between sites. In this form, we can explicitly calculate observables from Eq. \eref{eq_gpw}, which will serve as order parameters that govern the systems phase.

First, we can define an order parameter that acts as an effective measure for the correlation properties across the lattice as
\begin{equation} \label{eq_ord_p1}
\langle \hat{b}_{i} \rangle = \sum\limits_{n=0}^z \sqrt{n}f_{n}^{(i)} f_{n-1}^{*(i)} = \varphi_i.
\end{equation}
Next, the density properties of the system can be captured as
\begin{equation} \label{eq_ord_p2}
\langle \hat{n}_{i} \rangle = \sum\limits_{n=0}^z n | f_{n}^{(i)} |^2 = \rho_i.
\end{equation}
By using these relations,  Eq. \eref{eq_bhm} can be written as a sum of on-site contributions that depend on the previously defined order parameters
\begin{equation}
\hat{H} = \sum\limits_{i}^{N} \hat{H}_i,
\end{equation}
with each on-site Hamiltonian being given by
\begin{equation} \label{eq_locH}
\hat{H}_i = \frac{U}{2} \hat{n}_{i} (\hat{n}_{i} - 1) + (\epsilon_i - \mu) \hat{n}_{i} - \, J(\hat{b}_{i} + \hat{b}_{i}^\dagger)\sum\limits_{<i,j>} \varphi_j ,
\end{equation}
where $\varphi_i$ is taken to be real without loss of generality for the considered phases. In order to then determine the ground state of Eq. \eref{eq_locH}, we use the method of self-consistency for the local problems. Strictly speaking, $z$ should be infinite for a bosonic system. For numerical purposes, however, we limit the maximum $z$ to a sufficiently large cutoff to obtain convergence of the mean-field phases.

In the initial step of the self-consistent loop, we define a set of uniformly distributed random order parameters for each site. The local Hamiltonians are then diagonalised such that the global ground state can be identified. From this ground state, new order parameters are evaluated using Eq. \eref{eq_ord_p1} and Eq. \eref{eq_ord_p2} for each site. The loop continues with these new order parameters and the process is repeated until the ground state energy converges to a specified accuracy. At convergence, the mean-field ground state and phase of the system is known.

We finally note here that, within the context of mean-field theory, the way in which we will define phases necessarily differs from exact approaches. Rigorously, one should inspect correlation functions and winding numbers to check for the onset of superfluid order. Since we can not reliably define correlators and other such macroscopic order parameters in a site decoupled model, we instead work with local order parameters. As we will discuss later, however, the application of a percolation analysis can be used as an effective measure for the onset of true superfluidity and allows for the definition of a mean-field BG.

\section{Defining the Mean-Field Bose Glass} \label{sc_4}
In the homogeneous Bose-Hubbard model, we can directly characterise two phases -- the MI and SF -- based on whether $\varphi_i \, \forall \, i$ is zero or non-zero respectively. When we have a non-zero onsite potential present, it is possible to have a third intermediate phase to separate the insulating and superfluid domains, which is the BG. The methods we use here to determine different mean-field phases are based on percolation approaches used in the study of disordered phase transitions \cite{PhysRevLett.75.4075,PhysRevA.76.011602,PhysRevA.79.013623,Dell_Anna_2011,PhysRevB.85.020501,Niederle_2013,Nabi_2016,PhysRevA.98.023628}. In particular, we use the approach outlined by Ref. \cite{Niederle_2013}, and analyse percolations of superfluid clusters on a discrete function $S$ to define the three phases of the inhomogeneous Bose-Hubbard model. We label sites with integer fillings of density $\rho_i$ as MI sites with $S_i=0$, and non-integer as SF sites with $S_i=1$. We also define an additional function $F$ in order to determine the onset of the mean-field BG phase, i.e the presence of correlations. For this function, we have $F_i = 1$ for sites with finite $\varphi_i$, and $F_i = 0$ otherwise.

First, the standard MI can be characterised by a uniform, zero $S$ and $F$ across all lattice sites, with no transport. As the tunnelling becomes more significant, a BG phase is expected to appear, which is instead defined by some finite, localised transport properties. In other words, clusters of SF sites will begin to form, but do not percolate across the entire system. These clusters will of course have some finite correlations associated to them, captured by the $\varphi$ order parameters. Furthermore, the phase will have some finite mean-field correlation fraction $\mathcal{F}$, which we define as
\begin{equation}
\mathcal{F} = \frac{N_{\varphi}}{N_{sites}},
\end{equation}
where $N_{\varphi}$ is the total number of sites with finite $\varphi_i$ and $N_{sites}$ is the total number of sites. As the tunnelling is further increased, macroscopic phase coherence is expected to appear, giving rise to the SF phase. Unlike the BG phase, the SF has at least one percolating cluster of SF sites appearing in the discrete function $S$, allowing for a percolation probability $\mathcal{P}$ to be defined as
\begin{equation} \label{eq_PPerc}
\mathcal{P} = \frac{N_{span}}{N_{\varphi}},
\end{equation}
where $N_{span}$ is the number of SF sites in a percolating cluster. The phases and above relations for each case are summarised in Table \ref{table_pm1}. 

\begin{table}[t!]
	\caption{Phases of the inhomogeneous Bose-Hubbard model}
	\label{table_pm1}
	\centering
	\begin{tabular}{l c c} \\
		\hline\hline \\
		Phase$\,\,\,\,$ & $\mathcal{F} \,\,\,\,$ & $\mathcal{P}$ \\ \\ [0.2ex]
		
		\hline \\
		
		\textit{MI}$\,\,\,\,$ & $ = 0 \,\,\,\,$ & $ = 0 $ \\
		\textit{BG}$\,\,\,\,$ & $ > 0 \,\,\,\,$ & $ = 0 $ \\
		\textit{SF}$\,\,\,\,$ & $ > 0 \,\,\,\,$ & $ > 0 $ \\ [1.5ex]
		
		\hline
	\end{tabular}
\end{table}

Strictly speaking, the BG is defined as an insulating phase, which would imply vanishing $\varphi_i$ across the lattice. However, due to the presence of onsite fluctuations from the energy potential, there will always be sites with finite $\varphi_i$ in the mean-field BG, with each site essentially having its own MI to SF critical point. Indeed, the corresponding average order parameter $\bar{\varphi}$ may then fall below some threshold and be sufficiently close to zero, allowing for the phase to physically be viewed as macroscopically insulating. 

We also remark that these definitions differ from other mean-field methods \cite{Bissbort_2009,PhysRevA.81.063643,PhysRevA.83.013605}, which usually rely solely on the average order parameter $\bar{\varphi}$ and average local compressibility $\bar{\kappa}$ to distinguish different phases. With these, the MI and SF can be defined by $(\bar{\varphi}, \, \bar{\kappa}) \approx (0, \, 0)$ and $(\bar{\varphi}, \, \bar{\kappa}) \neq (0, \, 0)$ respectively.  The mean-field BG now has $\bar{\varphi} \approx 0$ and $\bar{\kappa} \neq 0$. However, as discussed in Ref. \cite{Niederle_2013}, the use of averaged order parameters leads to an exaggeration in critical points for finite disorder strengths, and will fail to predict an intermediate BG phase between a MI and SF phase in certain scenarios \cite{PhysRevLett.103.140402}.

By instead employing a percolation analysis, we can find signatures of global phase coherence in the system, with results from crystalline disordered systems in good agreement to those of quantum Monte-Carlo \cite{PhysRevLett.107.185301,PhysRevLett.103.140402,PhysRevLett.99.050403,pollet2013review} and tensor network \cite{Goldsborough_2015,Rapsch_1999} approaches. This is because the presence of energy modulations induces pronounced fluctuations to onsite particle numbers for finite $J/U$, giving rise to non-integer densities within SF clusters. The percolation of these clusters is intrinsically linked to the SF density calculated in quantum Monte-Carlo simulations, and is hence a very reliable measure for the onset of true superfluidity, even in the mean-field limit \cite{Niederle_2013}.


\section{Mean-Field Bose Glass} \label{sc_5}

\subsection{Characterising the Phase Transitions}
\label{sec:Character}

Due to the continuous behaviour of the local order parameters we consider here, we must define threshold conditions for the phases so that we can label different regions accordingly. We will use similar thresholds to those in Ref. \cite{Niederle_2013}, but we will motivate this choice in this section by investigating the transition points themselves. To check for integer densities, we evaluate the difference of $\rho_i$ with the nearest integer for each site. If this difference is less than $5 \times 10^{-3}$, then the density is considered integer, hence $S_i = 0$, otherwise it is non-integer and $S_i = 1$. In a similar manner, to define $F$, we look for sites with $\varphi_i > 10^{-2}$, i.e sites with finite mean-field correlations.

\begin{figure}
	\centering
	\includegraphics[width=0.98\linewidth]{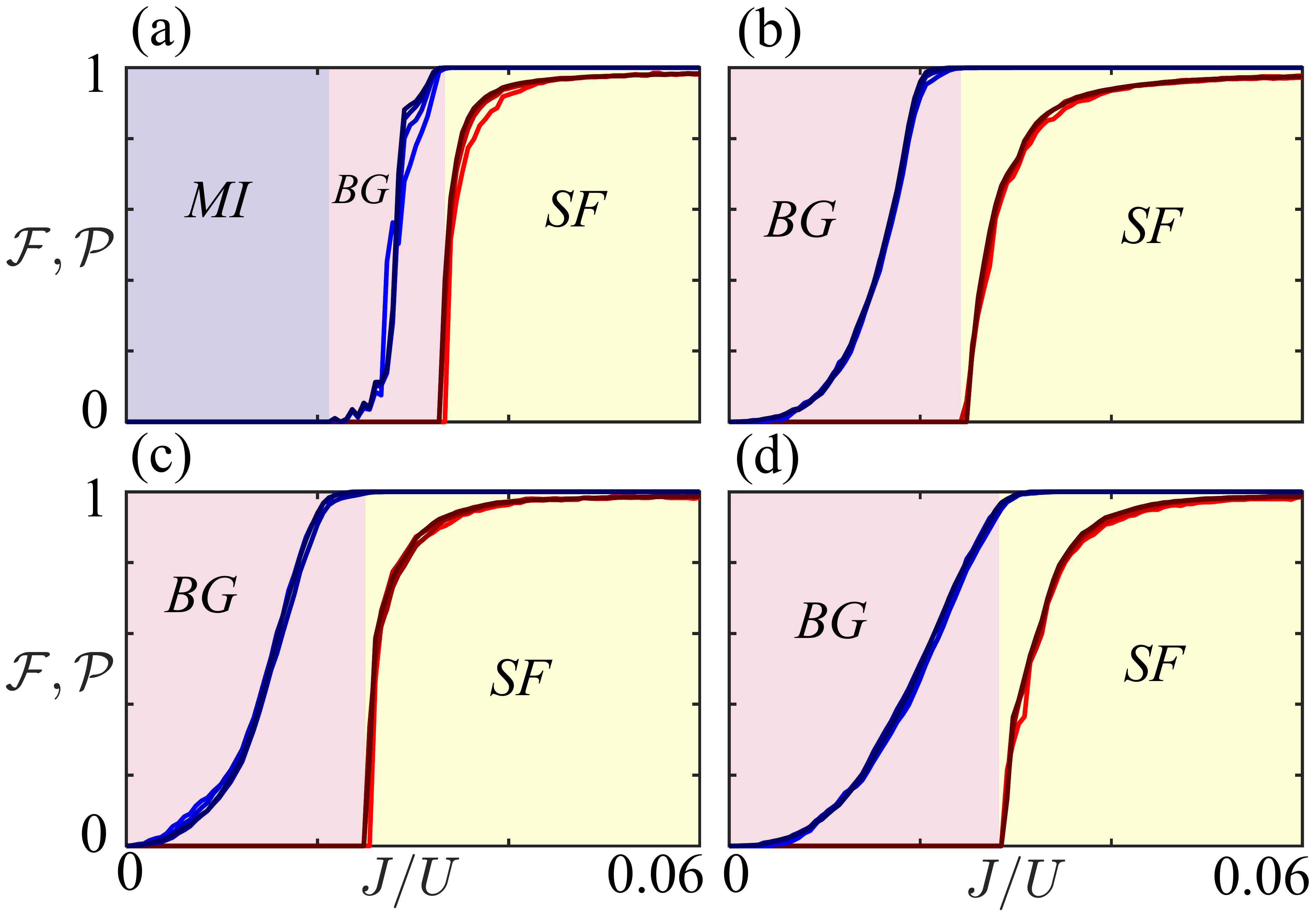}
	\caption{Plots of the mean-field correlation fraction $\mathcal{F}$ (blue curves) and percolation probability $\mathcal{P}$ (red curves) for system sizes of $N=33\times33$, $N=55\times55$, $N=77\times77$ and $N=99\times99$. The darker curves correspond to the larger system sizes. For each row, we fix the modulation strength to (a,b) $\lambda/U=0.15$ and (c,d) $\lambda/U=0.525$, whereas each column corresponds to a fixed chemical potential of (a,c) $\mu/U=0.6$ and (b,d) $\mu/U=0.2$. }
	\label{figure_ordC}
\end{figure}

We evaluate $\mathcal{P}$ by looking at the maximum extent of each SF cluster. If this extent is greater than $\beta K$, where $K$ is the maximum size of the lattice and $\beta = 0.99$ is a numerical scaling prefactor, then $\mathcal{P}$ takes a finite value as defined by Eq. \eref{eq_PPerc}. Otherwise, $\mathcal{P}$ will be zero if there is no percolation of the clusters. In the case of a square lattice, $K$ is simply the side length of the square. For the quasicrystalline tilings, $K$ is defined by $2R_m$, where $R_m$ is the maximum radial coordinate of the tiling with respect to the origin (central site).

We show an example slice of the order parameters in Fig.~\ref{figure_ordC} for the 2D AA model.  First,  it can be clearly observed that the mean-field BG to SF transition is relatively sharp. This reflects the enhanced definition of the mean-field BG via the percolation analysis.  Therefore,  any variations in the thresholds considered for the BG to SF transition point will result in little change to the critical points considered in this work.

The MI to BG transition is not as well defined for the mean-field approach, as the percolation analysis does not give this transition point.  This can be seen in the slower transition exhibited by $\mathcal{F}$ in Fig.~\ref{figure_ordC}.  Strictly speaking,  the MI phase is destroyed as soon as a single site exhibits non-zero transport.  However, this is not a practical definition due to the influence of finite size effects and numerical noise.  This requires the setting of a numerically non-zero but effectively zero threshold for the $\varphi$ order parameters in the definition of $\mathcal{F}$, as discussed earlier.  However,  due to the slowly increasing nature of $\mathcal{F}$ observed in Fig.~\ref{figure_ordC},  any changes in the threshold conditions can result in moderate changes in the MI to mean-field BG transition points.  These changes will never be substantial enough to remove the mean-field BG from the phase diagram of the quasicrystalline models considered here,  as $\mathcal{F} \sim 1$ before $\mathcal{P} > 0$.  From considering various cuts in constant $\mu/U$,  we have determined that a lattice size of $\sim 5000$ sites is sufficient to obtain convergence in the mean-field critical points to an average order of $10^{-4}$ for the BG to SF transition and $10^{-3}$ for the MI to BG transition. The convergence of the transition points with system size is shown in Fig.~\ref{figure_ordC} by the inclusion of four different system sizes, and we will consider full mean-field phase diagrams for different lattice sizes in Sec.~\ref{sec:AA}.

\subsection{Interacting 2D Aubry-Andr\'e Model}
\label{sec:AA}

\begin{figure*}[t]
	\centering
	\includegraphics[width=0.98\linewidth]{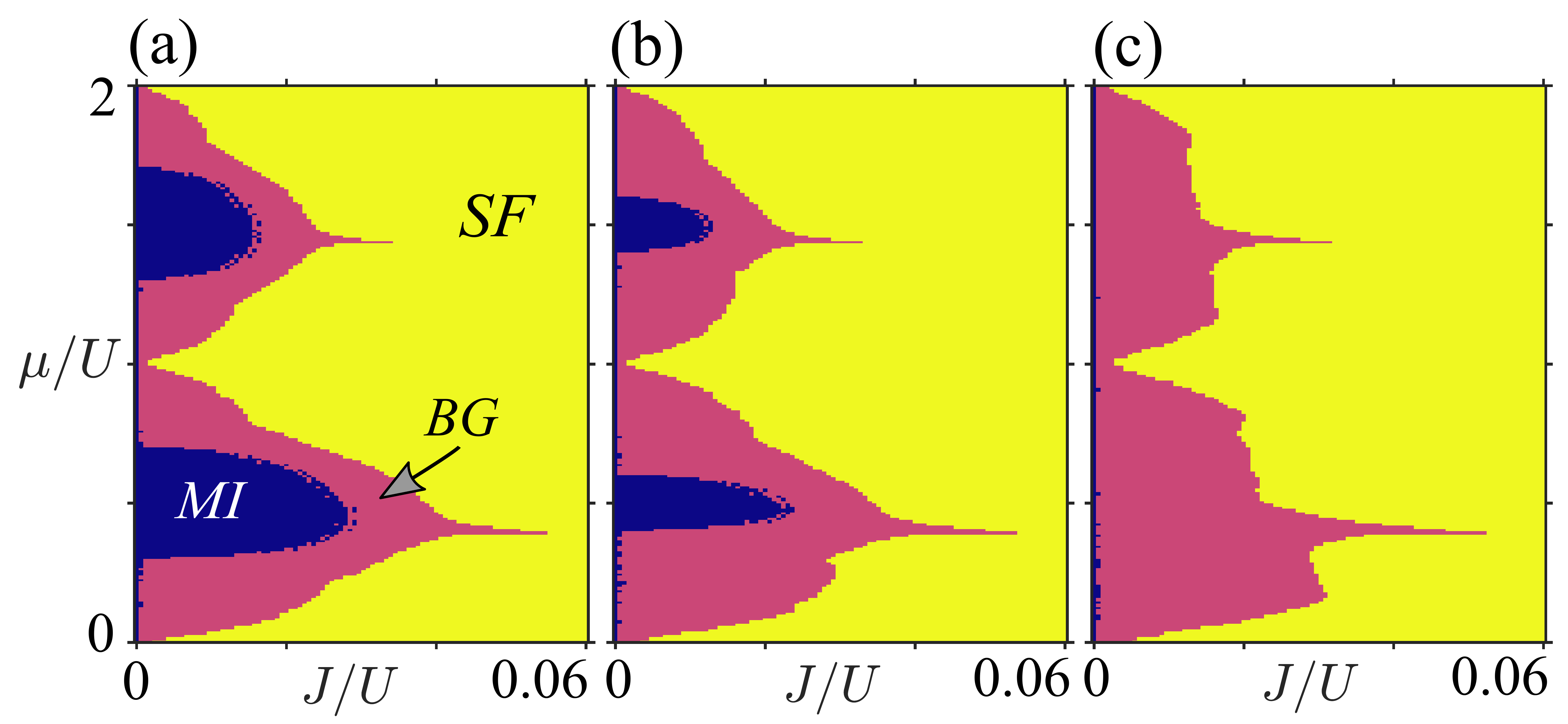}
	\caption{Phase diagrams of the 2D AA model for fixed modulation strengths $\lambda/U$. Here, we consider (a) $\lambda/U=0.15$, (b) $\lambda/U=0.20$ and (c) $\lambda/U=0.35$. As we increase $\lambda/U$, the MI lobes are slowly reduced in extent, leaving behind larger regions of the BG phase.}
	\label{figure_x8}
\end{figure*}

We will now turn our attention to the mean-field phase diagrams of the 2D AA model.  To reiterate, this model has a crystalline lattice geometry  (that of a square) and an onsite potential which is quasicrystalline and given by Eq. \eref{eq_aaEps}.  In Fig.~\ref{figure_x8} we show the mean-field phase diagrams for a $99 \times 99$ lattice,  which is sufficient for good convergence of the transition points as analysed in Sec.~\ref{sec:Character}.  As can be seen from these results, we find that the mean-field BG phase clearly separates the MI and SF domains,  as expected. By increasing $\lambda/U$, the mean-field BG slowly reduces the extent of the MI phase in Fig. \ref{figure_x8}(b) and \ref{figure_x8}(c), with the BG forming lobe-like structures, similar to the MI in a homogenous system. While the behaviour we observe here is comparable to what is seen in other inhomogeneous and disordered models \cite{Niederle_2013,PhysRevA.76.011602,PhysRevA.79.013623}, we also find several key differences between both results as well.

\begin{figure}[t]
	\centering
	\includegraphics[width=0.9\linewidth]{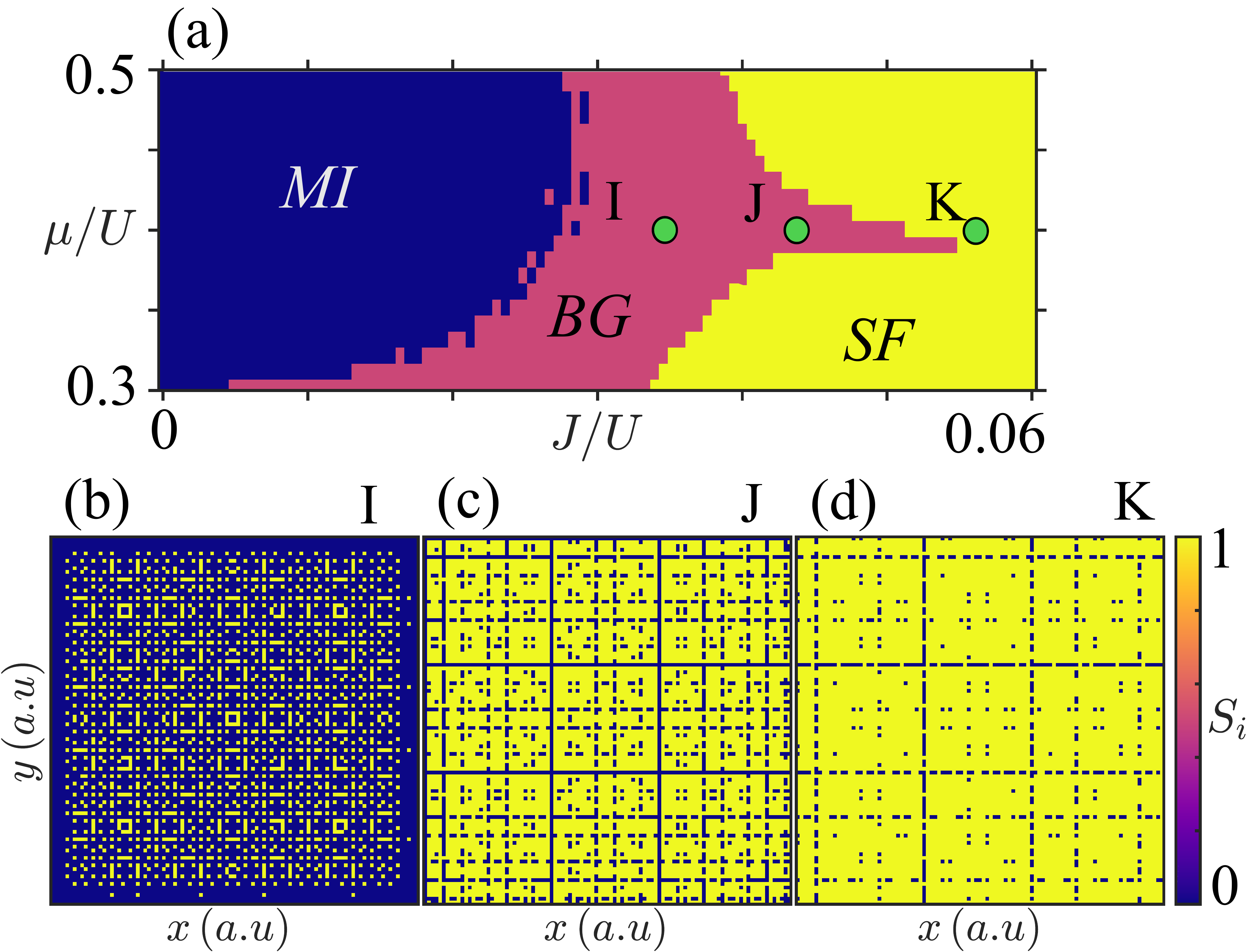}
	\caption{Plots of the (a) phase diagram in a reduced interval and (b-d) discrete functions $S$ on a line across the extruding feature with $\mu/U=0.39$ for $\lambda/U=0.15$. The labelled green circles in (a) have their corresponding discrete functions $S$ plotted for (b) $J/U=0.035$, (c) $J/U=0.043$ and (d) $J/U=0.056$. As we increase $J/U$, lines of weak modulation throughout the system prevent true percolation until a critical tunnelling rate.}
	\label{figure_x10}
\end{figure}

First, we find that extruding features can emerge from the mean-field BG lobes at different $\mu/U$, as observed in all three phase diagrams of Fig.~\ref{figure_x8}.  We plot the discrete function $S_i$,  which is non-zero if a site is in a SF state,  across one of these features in Fig. \ref{figure_x10} with fixed $\mu/U = 0.39$ and $\lambda/U=0.15$.  The three example states in Fig.~\ref{figure_x10} thus reflect moving further along the extrusion of the mean-field BG. From these figures, the reason for this curious feature on the phase diagram becomes immediately clear. The weak modulation lines discussed in Sec.~\ref{sec:InterAA} prevent the percolation of a macroscopic SF. This is shown clearly near the limits of the extruding feature in Fig.~\ref{figure_x10}(b),  where the state is a set of large SF domains, but no single cluster percolates across the full system, resulting in a weak BG phase.  Furthermore, the BG has a more intricate and self-similar structure near the start of this extrusion as shown in Fig.~\ref{figure_x10}(a),  with the state being constructed of many self-similar SF structures.  The extrusion of the mean-field BG is truly an artefact of the quasicrystalline potential, as such weak modulation lines would not exist if the potential was fully disordered. 

\begin{figure}[t]
	\centering
	\includegraphics[width=0.9\linewidth]{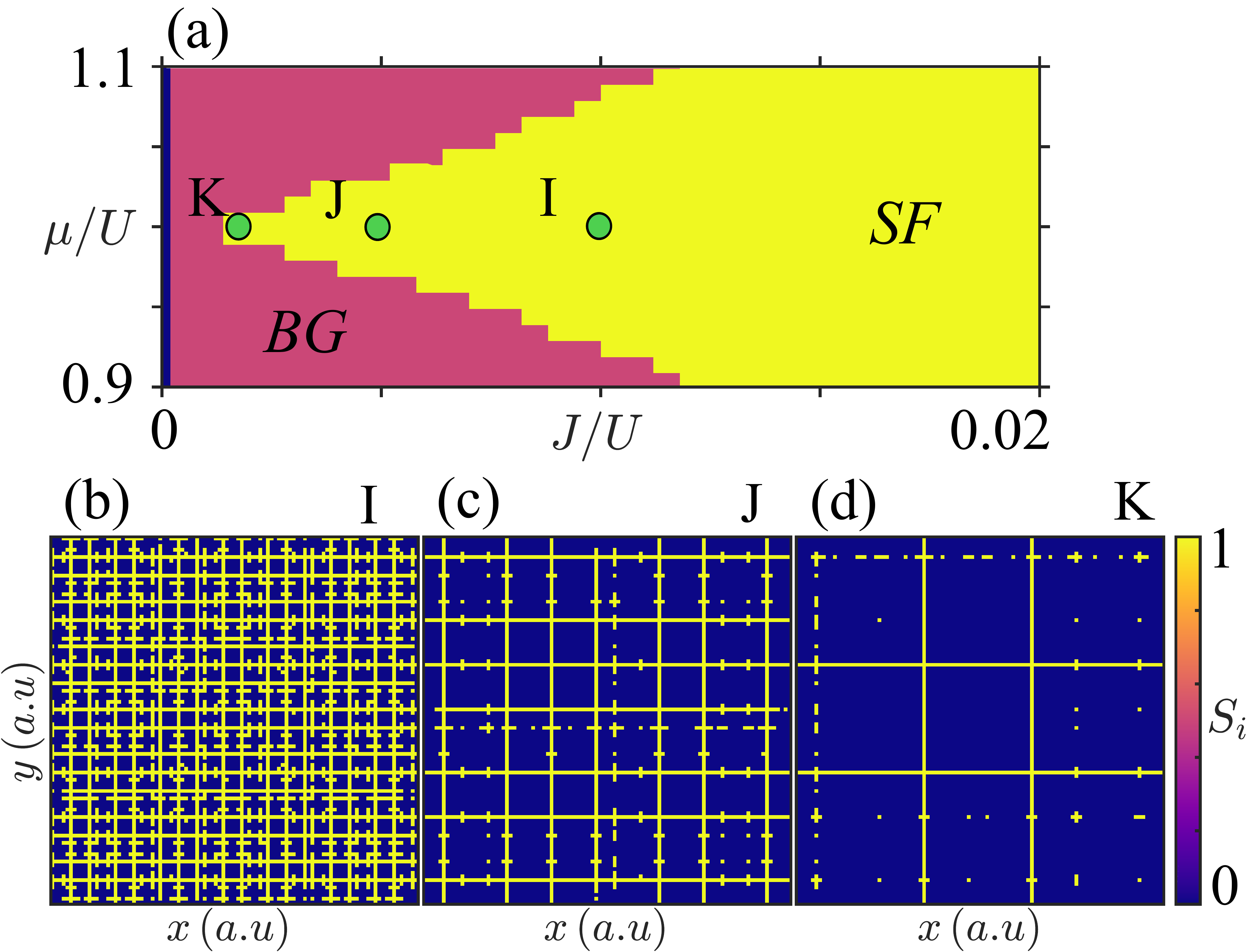}
	\caption{Plots of the (a) phase diagram in a reduced interval and (b-d) discrete functions $S$ to show the SF persistence at $\mu/U=1$ for $\lambda/U=0.2$. The labelled green circles in (a) have their corresponding discrete functions $S$ plotted for (b) $J/U=0.01$, (c) $J/U=0.005$ and (d) $J/U=0.002$. At small $J/U$, lines of weak modulation allow for percolation to occur, which stabilises the SF phase.}
	\label{figure_x12}
\end{figure}

In normal disordered realisations of the BG, the phase will replace MI lobes and be present across the full range of $\mu/U$,  thus it lacks the lobe structure that is associated to the MI domain \cite{Fisher1989}.  Furthermore, intermediate disorder strengths will decrease the width of the MI lobes across $\mu/U$ \cite{Niederle_2013,PhysRevA.76.011602,PhysRevA.79.013623}.  Interestingly,  with a quasicrystalline potential,  the mean-field BG retains a lobe-like structure even for large modulation strength $\lambda/U$,  as shown in Fig.~\ref{figure_x8}.  However, we do observe the decrease in width of the MI phase normally associated with the appearance of the BG state. 

To investigate the retention of the lobe-like structure,  we again consider discrete functions $S_i$ at fixed $\mu/U$.  In this case, the region of interest is at integer chemical potentials, where the SF phase is stabilised to allow for the lobe structure to form.  We show a variety of $S_i$ for small $J/U$ at $\mu/U = 1$ with $\lambda/U = 0.2$ in Fig.~\ref{figure_x12}.  The reason for the SF stabilisation and BG suppression can again be linked to the weak modulation lines. Unlike the previous feature in Fig.~\ref{figure_x10} where the BG was extended due to lines of insulators, the BG is now suppressed at integer $\mu/U$ due to the SF phase percolating across lines of weak modulation. Similar behaviour is also observed at other $\lambda/U$ and integer $\mu/U$.

\begin{figure*}
	\centering
	\includegraphics[width=0.98\linewidth]{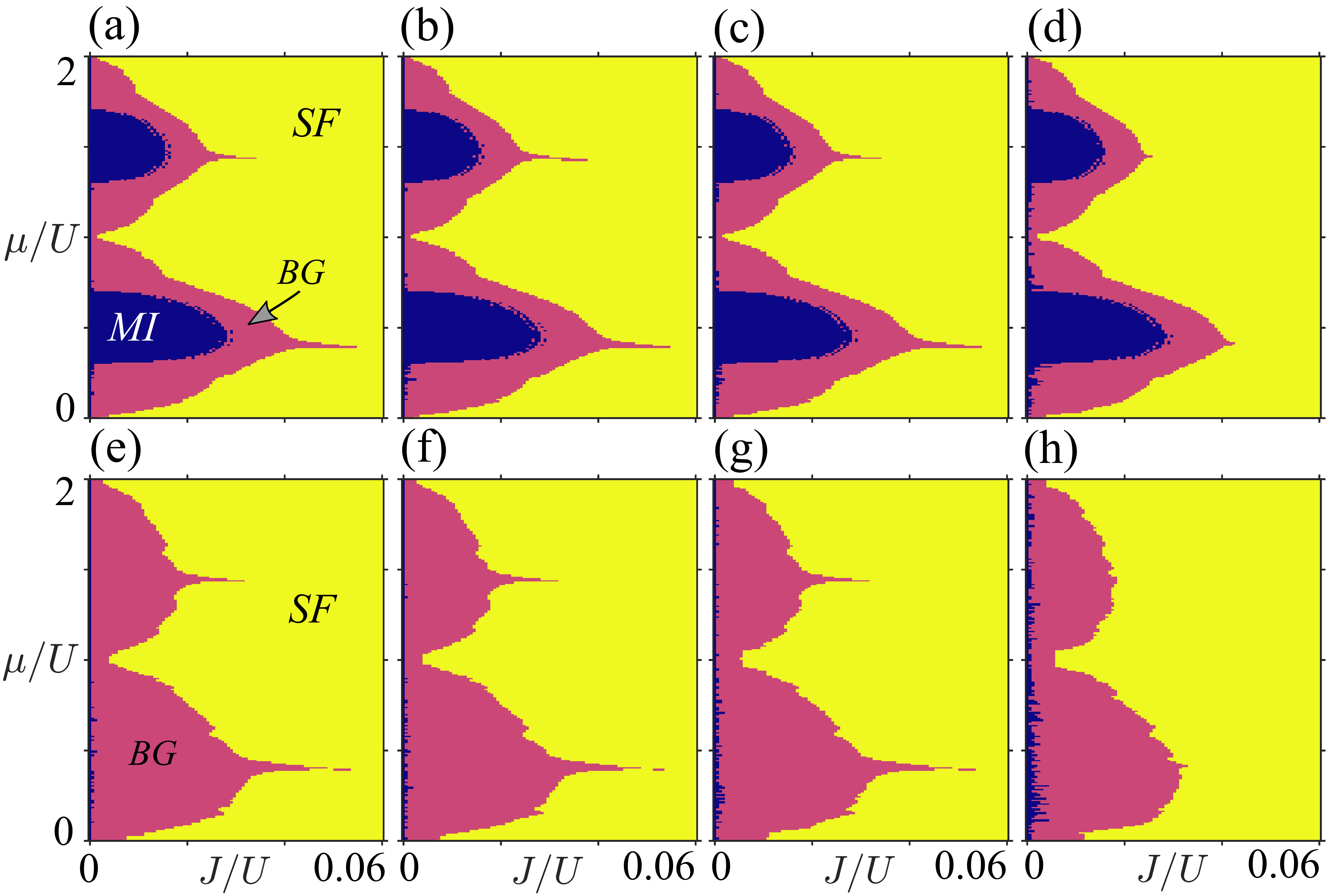}
	\caption{Phase diagrams of the 2D AA model for fixed modulation strengths (a-d) $\lambda/U=0.15$ and (e-h) $\lambda/U=0.525$. Each column corresponds to a system size of (a,e) $N=99\times99$, (b,f) $N=77\times77$, (c,g) $N=55\times55$ and (d,h) $N=33\times33$.}
	\label{figure_x13}
\end{figure*}

An interesting question is then, how does the appearance of weakly modulated lines change global properties of the phase diagrams for different system sizes? To answer this, we look at two specific $\lambda$, one relatively moderate and one large,  and plot the corresponding mean-field phase diagrams in Fig.~\ref{figure_x13} for system sizes between $33\times33$ and $99\times99$.  We can immediately see that the mean-field critical points have very little variation within this range of system sizes, and the majority of features are invariant against changes to the system size. This is to be expected from the previous analysis in Sec.~\ref{sec:Character}. However, for the smallest of lattices,  i.e.  $33\times33$, the features corresponding to the weak modulation lines are less pronounced.  This is again not surprising due to the smaller number of sites in general and,  hence, fewer regions in the quasicrystalline potential that will exhibit the behaviour of weak modulation lines. We also note that this is reflective of other quasicrystalline properties that can often be dependent on the system being of a sufficient size \cite{Loring2019,duncan2019topological}. In other words, the partial dependence on the system size is a result of the long-range order present.

\subsection{Weak Modulation Lines}

\begin{figure}[t]
	\centering
	\includegraphics[width=0.8\linewidth]{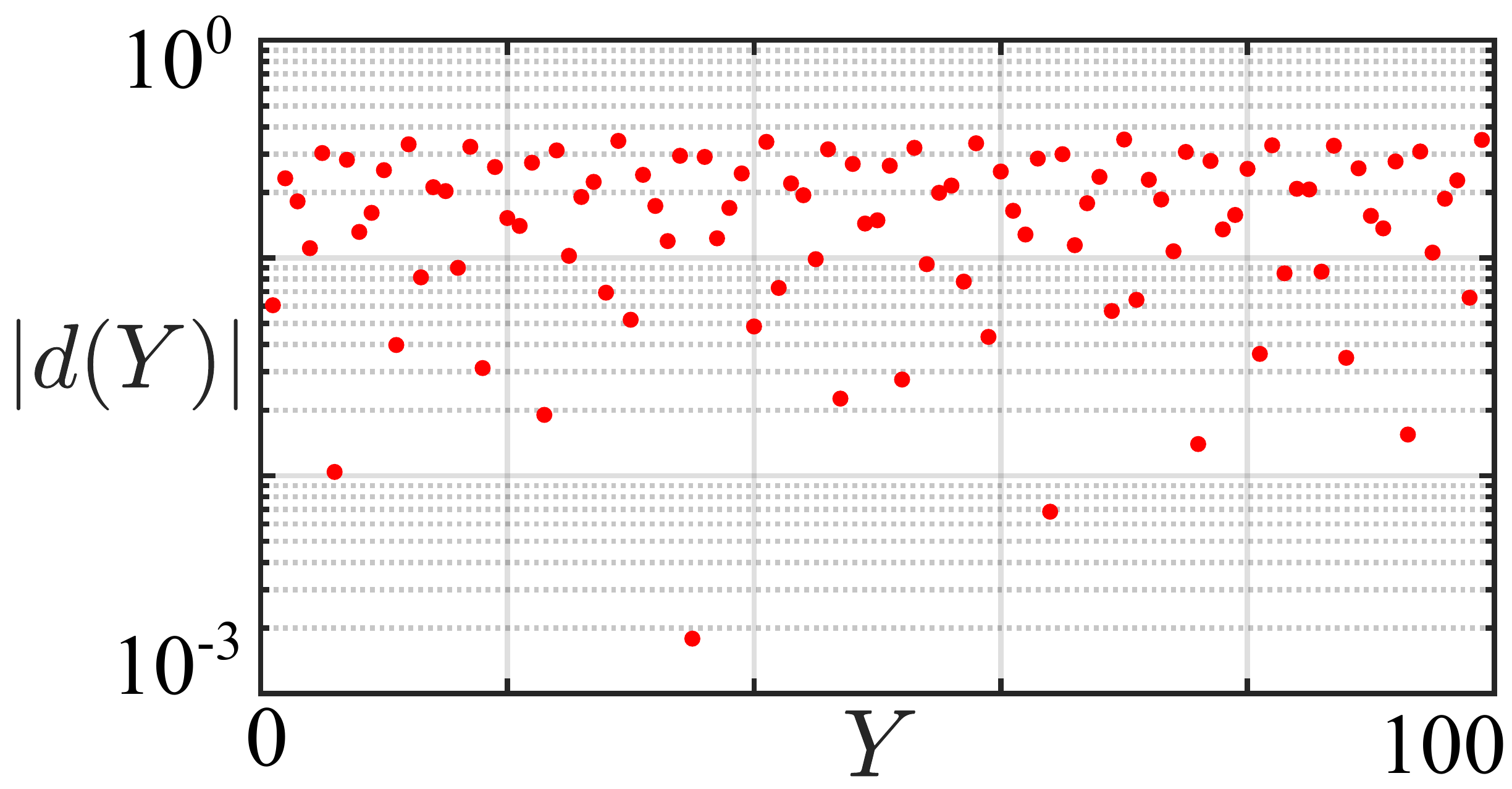}
	\caption{Plots of the difference function $d(Y)$ for the first 99 $Y$ columns. Throughout this range of $Y$, several points will minimise the difference function to less than $10^{-2}$, which directly correspond to the columns of weak modulation appearing in the energy distribution.}
	\label{figure_disorderLn}
\end{figure}

Our results have shown that the mean-field BG does appear in the presence of a quasicrystalline potential.  The quasicrystalline nature results in the stabilisation and suppression of the mean-field BG in different scenarios. To better understand the origin of the weak modulation lines, we now turn our attention to how they arise in the potential.  First, we rewrite Eq. \eref{eq_aaEps} as
\begin{equation} 	\label{eq_aaEps2}
\epsilon_i = -2 \lambda \cos(2 \pi \beta x_i) \cos(2 \pi \beta y_i).
\end{equation}
We know from the form of the weak modulation lines that they can appear in single rows or columns.  By then fixing one position index $y_i = Y$, the appearance of a weakly modulated line where $\epsilon_i = 0$ can be defined as
\begin{equation}
2 \pi \beta Y = \frac{k \pi}{2},
\end{equation}
where $k$ is an odd integer. In other words, if
\begin{equation}	\label{eq_aaEps_c}
d(Y) = Y - \frac{k}{4 \beta} = 0,
\end{equation}
then the corresponding set of sites $(x_i,Y)$ will be a weakly modulated line with $\epsilon_i = 0$.  Note,  we could have fixed $x_i$ in the beginning and obtained the same condition for a set of points with fixed $x$,  meaning the derivation confirms the presence of weakly modulated lines in both dimensions parallel to the boundary of the lattice. We could also fix a relationship between $x_i$ and $y_i$ which would allow for the same derivation to show that general weak modulation lines are possible, e.g. at some angle to the boundary of the lattice. Similar weak modulation lines could appear in other potentials of a quasicrystalline nature.

Taking $\beta=1/\sqrt{2}$, we plot the function $d(Y)$ in Fig. \ref{figure_disorderLn} by considering the difference of each $Y$ of the lattice with the closest point in the set $k/4\beta$, with $k$ an odd integer.  In practice the condition given by Eq. \eref{eq_aaEps_c} can be relaxed to allow for weak modulation lines to effectively occur whenever
\begin{equation}
d(Y) = Y - \frac{k}{4 \beta} \approx 0,
\end{equation}
which allows some small fluctuation along the lines due to the cosine potential,  but these modulations would be weak.  Here, we see that there are indeed a few specific $Y$ that meet the condition $d(Y) \approx 0$,  which directly correspond to the weakly modulated lines observed in Figs. \ref{figure_x10} and \ref{figure_x12}. Therefore,  the weakly modulated lines are predictable from the form of the potential, and any other quasicrystalline potentials could be checked to see if such lines exist prior to any calculations.  From Fig.~\ref{figure_disorderLn},  it is also clear that if the system is not of sufficient size, then there will be fewer weak modulation lines to stabilise/destabilise the mean-field BG,  as $d(Y) \approx 0$ will be satisfied less often.  This predicts the behaviour exhibited for the smallest system shown in Fig.~\ref{figure_x13},  where the weak modulation line features are less pronounced.
 
If weak modulation lines are present in the potential, then we would expect the same physics to be observed for the mean-field BG as has been shown in Sec.~\ref{sec:InterAA}. This includes the stabilisation of the mean-field BG around non-integer $\mu/U$ due to insulating lines in Fig.~\ref{figure_x10}, and the suppression of the BG due to SF percolation across the same lines in Fig.~\ref{figure_x12}.

\subsection{Disordered Vertex Model}

We have so far focused on the quasicrystalline nature of the problem being present in the onsite potential of the model.  In considering the disordered vertex model,  we now turn to a different scenario, where the onsite potential is disordered and the underlying lattice geometry is that of a quasicrystal. The lattice is constructed as discussed in Sec.~\ref{sec:DisVert} and illustrated in Fig.~\ref{figure_x3}, with the Hamiltonian still given by Eq. \eref{eq_bhm}, but now with a uniform random disorder in the onsite potential.  Due to the presence of random disorder, multiple mean-field critical points are calculated and averaged over for different random initialisations of the onsite potential. We will consider $250$ disorder realisations for each critical point,  with the error being sufficiently low from this choice. All critical points are plotted with error bars from the analysis of the disorder realisations, however, the errors are of the order of the point size plotted.

\begin{figure*}
	\centering
	\makebox[0pt]{\includegraphics[width=0.98\textwidth]{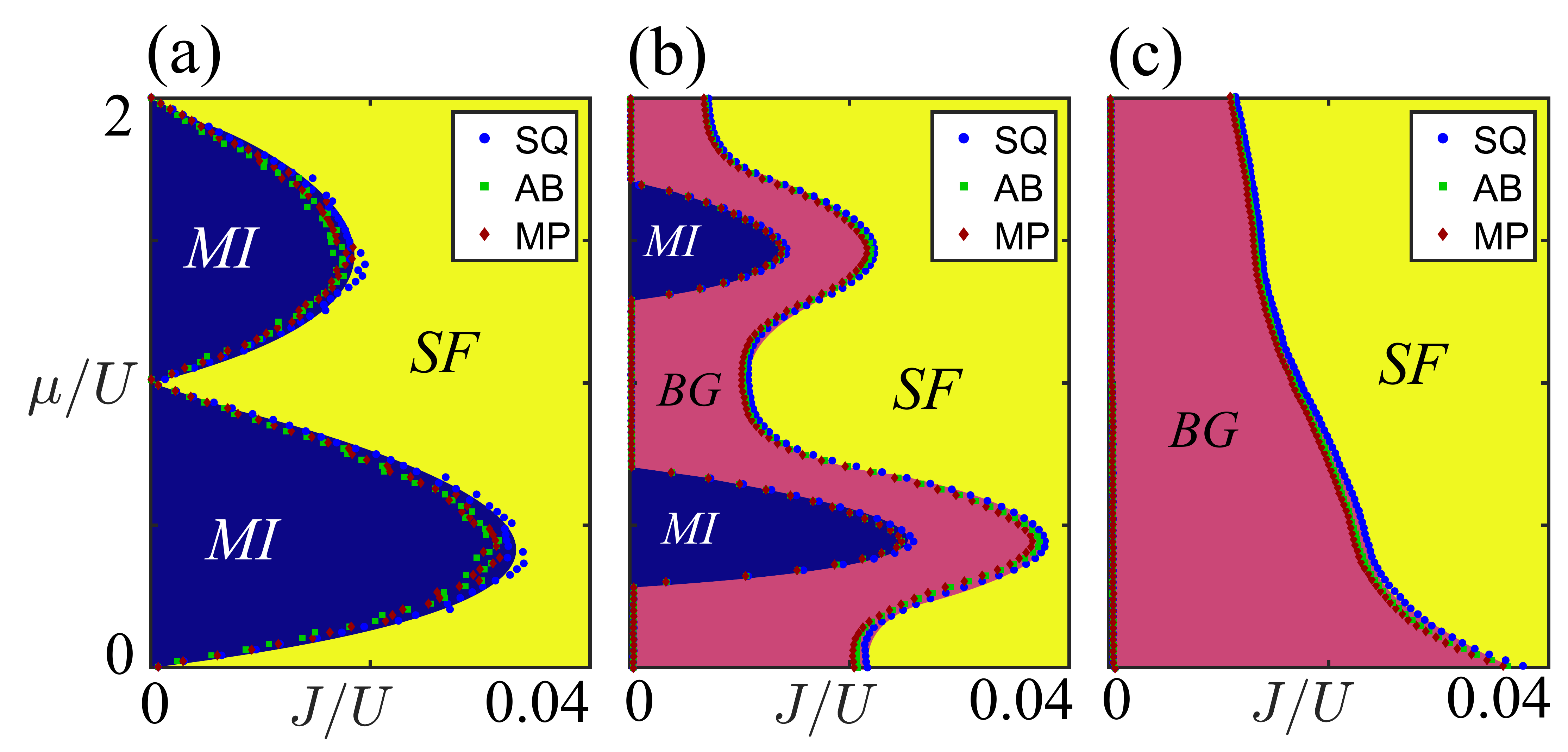}}
	\caption{Phase diagrams of the disordered Bose-Hubbard model for (a) $\Delta/U = 0.0$, (b) $\Delta/U = 0.6$ and (c) $\Delta/U = 2.0$. The coloured points indicate the numerically obtained critical points, with blue circles, green squares and red diamonds denoting the square (SQ), Annmann (AB) and Penrose (MP) lattices respectively. Based on these points, the phase regions are marked accordingly. The well-known Mott lobe structures are observed in the disorder free scenario when $\Delta/U = 0.0$. As we enlarge $\Delta/U$, we find that the BG phase clearly separates SF and MI domains, before enveloping the MI completely in the strong disorder limit.}
	\label{figure_x5}
\end{figure*} 

For comparison, we will consider an 8-fold Ammann-Beenker (AB) quasicrystal,  a 5-fold Moore-Penrose (MP) quasicrystal, and a square (SQ) lattice.  All three lattices have $\sim 3000$ sites,  with the AB having $3041$,  the MP having $3056$, and the square having $3025$ sites. Note, the rhombic tilings of the AB and MP lattice result in vertex models with average coordination number in the bulk $\approx 4$, meaning that the square lattice is a good crystalline comparison. The mean-field phase diagrams for all three of these models across a range of random disorder strengths are shown in Fig.~\ref{figure_x5}.  We observe a surprisingly good convergence of the mean-field critical points for all three models. In addition, we do not observe the suppression or stabilisation of the mean-field BG due to the quasicrystalline structure.

It is to be expected that the mean-field phase diagrams would converge between periodic and quasicrystalline lattices with no random onsite disorder as shown in Fig.~\ref{figure_x5}(a) \cite{PhysRevA.100.053609,ghadimi2020mean}. When random disorder is included, we observe the same features as expected from previous results on disorder in periodic lattices \cite{Fisher1989}.  The results also show that there is no direct MI to SF transition in the presence of random disorder in a quasicrystal,  as is the case in crystalline lattices \cite{PhysRevB.80.214519,PhysRevLett.107.185301,PhysRevLett.103.140402,PhysRevLett.99.050403,Goldsborough_2015,Rapsch_1999}.  Small fluctuations in the critical points of the AB, MP, and square lattices are a result of the varying coordination number in the quasicrystals and the difference in the rotational symmetry that must be exhibited by the ground state, i.e. 8-fold, 5-fold, and 4-fold.

\section{Conclusions} \label{sc_6}

We have confirmed the presence of a mean-field BG phase in quasicrystalline models using the Gutzwiller mean-field method for a Bose-Hubbard Hamiltonian. The characterisation of the MI to BG transition within the mean-field approach is introduced by studying the correlation fraction of the state, and suitable thresholds investigated to consistently define the critical point. On the other hand, the BG to SF transition was identified through the use of a percolation analysis, which allows for a consistent definition of a BG like phase in the mean-field.

Two different quasicrystalline models were considered.  First, the 2D AA model,  which has an underlying periodic square lattice geometry but a quasicrystalline onsite potential. Second, the disordered vertex model, which has a quasicrystalline lattice geometry with a random uniform disorder in the onsite potential. Both these models contain a quasiperiodic component, however, we observe that the different areas in which they manifest is of vital importance.  In both models we confirm the presence of a mean-field BG phase, which is an intermediary for the macroscopic MI and SF. 

The disordered vertex models show effectively no difference between periodic and aperiodic lattices. The critical points for the MI to SF,  MI to BG, and BG to SF show only marginal variation between quasicrystalline and crystalline models. The small fluctuations exhibited in the critical points are accounted for by the non-constant coordination number for the quasicrystals and the differing rotational symmetries that the ground state must exhibit.  As all rhombic aperiodic tilings have an average coordination number of $\approx 4$, we would expect the mean-field results of this model to hold for other rhombic tilings as well.  Furthermore, we would also expect that the critical points will scale with the average coordination number for any disordered vertex model.  This would be true as long as the distribution of coordination numbers does not become large and uniform, which would result in a different critical point.  However, we note here that we do not know of any aperiodic tiling that would result in such a distribution of the coordination number.  

For the 2D AA model, we confirm that the mean-field BG phase is exhibited by periodic lattices with quasiperiodic potentials.  In this case, the mean field BG phase is stabilised or suppressed in different scenarios due to the presence of weak modulation lines.  We showed that the weak modulation lines can be predicted from a simple analysis of the quasiperiodic potential. This could be useful for their prediction in other models outside of the mean-field analysis considered in this work.  For non-integer $\mu/U$, the weak modulation lines can result in a stabilisation of the BG phase over a larger range of $J/U$ than expected,  giving the protruding features of the mean-field BG shown in Fig.~\ref{figure_x8} and \ref{figure_x13}.  We showed that this feature is due to the weak modulation lines blocking the formation of a macroscopic SF state. This can result in large regions of SF (but not of the same size as the state) separated by thinner regions of MI. For integer $\mu/U$, we observe the opposite scenario. The mean-field BG is now suppressed due to the percolation of thin SF layers that form on weakly modulated lines resulting in a macroscopic SF. The combination of this suppression and enhancement at different $\mu/U$ conspires so that the mean-field BG gains a lobe-like structure, reminiscent of the ubiquitous MI lobes. This is significant, as the BG is usually formed across the majority of the small tunnelling domains; with little structure as a function of the chemical potential, as is exhibited in the disordered vertex model.  The weak modulation lines of the 2D AA model are a result of the quasicrystalline structure of the potential, and could be exhibited by other quasiperiodic potentials. This means that the lobe-like structure exhibited by the mean-field BG could be a signature that the BG phase is due to a quasicrystalline potential instead of a uniform random potential.

\textit{Comment} -- During the finalisation of this manuscript, a study considering optical lattice quasicrystals using the Gross-Pitaevskii equation confirmed the presence of the BG phase in the continuous regime \cite{Gautier2021}. In that work, it is shown that due to the inhomogeneous and quasiperiodic nature of the potential itself, a BG phase can indeed be stabilised in potentially experimentally relevant regimes. While the considered models and approaches are different, the results in this manuscript complement those in Ref.~\cite{Gautier2021}.

\ack
The authors thank Anne E. B. Nielsen and Andrew J. Daley for useful discussions. D.J., acknowledges support from EPSRC CM-CDT Grant No. EP/L015110/1. C.W.D. acknowledges support from the Independent Research Fund Denmark under Grant Number 8049-00074B. Work at the University of Strathclyde was supported by the EPSRC Programme Grant DesOEQ (EP/P009565/1), and by the European Union's Horizon 2020 research and innovation program under grant agreement No.~817482 PASQuanS.

\appendix
\section{Vertex Models for Quasicrystals} \label{sc_QC}
In order to generate the quasicrystalline tilings used in this work, we consider 2D cut-and-project sets of a $D$ dimensional lattice $\mathcal{Z}$, which provides a mapping from $\mathbb{R}^D \rightarrow \mathbb{R}^2$. For the following, we consider the $\mathcal{Z}$ to be hypercubic structures, in which the generators of $\mathcal{Z}$ are defined by permutations of $1$ within a null vector existing in $\mathbb{R}^D$. The first step to produce the tiling is to define a rotation on the points $\vec{V} \in \mathcal{Z}$ relative to the origin $\vec{0}$
\begin{equation} \label{eq_RMV}
\vec{W} = \mathcal{R}\vec{V},
\end{equation}
where $\vec{V}$ is specified by the generators of $\mathcal{Z}$, $\mathcal{R}$ is an incommensurate rotation operator and $\vec{W}$ is a transformed position. For hypercubic lattices, the columns of $\mathcal{R}$ naturally define the transformed generators, leading to the following constraints on $\mathcal{R}$
\begin{equation}
\mathbf{C}_i \cdot \mathbf{C}_j = \delta_{ij},
\end{equation}
where $\mathbf{C}_i$ is the ith column of $\mathcal{R}$ in vector form and $|\mathcal{R}|=1$ for a rotation operator. From the $\mathbb{R}^D$ superspace of $\mathcal{Z}$, we can define two unique subspaces as $\mathcal{T} \in \mathbb{R}^2$, the tiling space of $\mathcal{Z}$ and $\mathcal{I} \in \mathbb{R}^{D-2}$, the internal space of $\mathcal{Z}$. Both $\mathcal{T}$ and $\mathcal{I}$ are mutually orthogonal subspaces defined through a duality map. The space $\mathcal{T}$ is simply the 2D projection of the transformed points $\vec{W}$, with the points in $\mathcal{T}$ being defined by the first two elements of $\vec{W}$. Similarly, the next $D-2$ elements of $\vec{W}$ define a point in $\mathcal{I}$.

By inspection, simply projecting all $\vec{W}$ to $\mathcal{T}$ will result in a dense set points in $\mathbb{R}^2$, with no particular aperiodic structure. The idea is then to limit the projected set of points in $\mathcal{R}$ using some cutoff in $\mathcal{I}$. In most cases, this is defined through the unit cell of $\mathcal{Z}$, which we will label as $\Lambda$. We now take the points spanned by $\Lambda$, apply the previous transformation in Eq. \eref{eq_RMV} and project $\Lambda$ to $\mathcal{I}$. In forming the tiling pattern for $\mathcal{T}$, we now only accept points in $\mathcal{T}$ whose dual in $\mathcal{I}$ falls within the convex hull spanned by the projected $\Lambda$ to $\mathcal{I}$. By using the unit cell as the bounding volume in $\mathcal{I}$, we ensure that multiple tiles/edges do not overlap.

In our work, we have considered the AB and MP tilings, which are generated from 4D and 5D hypercubic lattices respectively.

\section{Rotation Matrices} \label{sc_AppRm}
Here, we give the precise forms of $\mathcal{R}$ that were used to construct the quasicrystalline tilings used in this paper.

\subsection{Ammann-Beenker Tiling}
The AB tiling can be generated as a canonical projection of a 4D hypercubic lattice $\mathcal{Z}$. A general position in $\mathcal{Z}$ can be written as
\begin{equation}
\vec{V} = \pmatrix{a_1 \cr a_2 \cr a_3 \cr a_4}
\end{equation}
where each $a_i \in \mathbb{Z}$. The rotation of $\vec{V}$ to $\vec{W}$ is defined by a $4 \times 4$ rotation matrix $\mathcal{R}$ and can be written as
\begin{equation}
\mathcal{R} = \frac{1}{2}
\pmatrix{
0 & 1 & \sqrt{2} & 1 \cr
\sqrt{2} & 1 & 0 & -1 \cr
\sqrt{2} & -1 & 0 & 1 \cr
0 & 1 & -\sqrt{2} & 1
},
\end{equation}

\subsection{Moore-Penrose Tiling}
The MP tiling can be generated as a canonical projection of a 5D hypercubic lattice. Similar to before, a general position can be expressed as
\begin{equation}
\vec{V} = \pmatrix{
a_1 \cr a_2 \cr a_3 \cr a_4 \cr a_5
}.
\end{equation}

$\mathcal{R}$ will now be a $5 \times 5$ rotation matrix, which can be written as
\begin{equation}
\mathcal{R} = \sqrt{\frac{2}{5}}
\pmatrix{
0 & s_2 & s_4 & -s_4 & -s_2 \cr
1 & c_2 & c_4 & c_4 & c_2 \cr
1 & c_4 & c_2 & c_2 & c_4 \cr
0 & s_4 & -s_2 & s_2 & -s_4 \cr
\sqrt{2}^{-1} & \sqrt{2}^{-1} & \sqrt{2}^{-1} & \sqrt{2}^{-1} & \sqrt{2}^{-1}
},
\end{equation}
where
\begin{equation}
s_n = \sin\Big(\frac{n\pi}{5}\Big),
\end{equation}
and
\begin{equation}
c_n = \cos\Big(\frac{n\pi}{5}\Big).
\end{equation}

\vspace{1cm}


\providecommand{\newblock}{}

\end{document}